\documentclass[preprint,showpacs,preprintnumbers,amsmath,amssymb]{revtex4}
\usepackage{dcolumn}
\usepackage{bm}
\usepackage{graphicx}%
\usepackage{epsfig}

\begin{document}

\title{Open system approach to the internal dynamics \\ of a model multilevel molecule}
\author{F. Giraldi \\{\footnotesize\it Quantum Research Group, School of Chemistry and Physics, University of KwaZulu-Natal \\ and National Institute for Theoretical Physics (NITheP), KwaZulu-Natal, Westville Campus, Durban, South Africa \\
Gruppo Nazionale per la Fisica Matematica, GNFM-INdAM, Via Madonna del Piano 10,I-50019 Sesto Fiorentino (FI), Italy\\
\& giraldi@ukzn.ac.za,figli@libero.it}\\[2ex] F. Petruccione {\\{\footnotesize\it Quantum Research Group, School of Chemistry and Physics, University of KwaZulu-Natal \\ and National Institute for Theoretical Physics (NITheP), KwaZulu-Natal, Westville Campus, Durban, South Africa \\\& petruccione@ukzn.ac.za}}}

\begin{abstract}
A model multilevel molecule described by two sets of rotational internal energy levels of different parity and degenerate ground states, coupled by a constant interaction, is considered, by assuming that the random collisions in a gas of identical molecules, provoke transitions between adjacent energy levels of the same parity. The prescriptions of the continuous time quantum random walk are applied to the single molecule, interpreted as an open quantum system, and the master equation driving its internal dynamics is built for a general distribution of the waiting times between two consecutive collisions.
 The coherence terms and the populations of the energy levels relax to the asymptotics with inverse power laws for relevant classes of non-Poissonian distributions of the collision times.
The stable asymptotic equilibrium configuration is independent of the distribution.
   The long time dynamics may be hindered by increasing the tail of the distribution density. This effect may be interpreted as the appearance of the quantum Zeno effect over long time scales.
 \end{abstract}

\maketitle

\section{Introduction}

The theory of Open Quantum Systems \cite{BP} provides a detailed description of the time evolution of an open system in terms of the quantum dynamical semigroups \cite{AL}. The Gorini-Kossakowski-Sudarshan-Lindblad form is the most general completely positive trace-preserving master equation mimicking the dynamics of the reduced density matrix describing the system of interest \cite{GKS76,Lindblad76}.
The prescription of the continuous time random walk, originally developed by Montroll and Weiss \cite{MW}, has been adopted by Budini \cite{CTQRW0,budiniPRE04} in order to build up a class of evolution equations, fulfilling the completely positive condition. The interaction of the open system with its external environment is interpreted as a random event and the elapsed time between two consecutive collisions represents a random renewal process. If the distribution of random time intervals between two consecutive jumps is non-Poissonian, anomalous diffusion is obtained from the random walk. This powerful technique, adapted to Open Quantum Systems, is named Continuous Time Quantum Random Walk (CTQRW) and leads to a convoluted structure master equation as a subordination \cite{S} to a Lindblad dynamics \cite{GKS76,Lindblad76}.

A gas of colliding identical model multilevel molecules has been considered in Ref. \cite{Zeno0}. The Rabi-like oscillations between the two sets of rotational internal energy levels of the molecule itself are shown to be inhibited by an increase of the mean collision time. This behavior has been interpreted as the appearance of the Quantum Zeno effect (QZE).
The QZE \cite{MS} is generally interpreted as the hindrance of the dynamics of an unstable quantum system, caused by frequent measurements. Experimental evidence of the QZE were shown by Cook \cite{Cook} in 1988, by Itano et al. \cite{IHBW} in forced Rabi oscillations between discrete atomic levels, and in spontaneous decaying systems by Fisher et al. \cite{FGR}, to name a few.
Under particular conditions the QZE is also obtained over short time scales \cite{HW,BN,K}, the literature on this argument is vast.

The model molecule adopted in Ref. \cite{Zeno0} refers to the experimental behavior of the nuclear spin depolarization in $^{13}CH_3F$ molecule \cite{CH3F}. The quantum number of the total spin of the three protons takes the values $3/2$ (orto) and $1/2$ (para), the transition between states of different parity are forbidden in the electric dipole interaction and the spin flip emerges from a weak coupling between two levels of different spin parity. Here, the freezing of the spin relaxation due to an increase of the gas pressure, is considered as the appearance of the dephasing caused by molecular collisions.
Also, inverse power law behavior emerges over long time scales in the time evolution of the survival probability of an unstable quantum system \cite{NNP}.

 In this scenario, we aim to study the internal dynamics of the model multilevel molecule adopted in Ref. \cite{Zeno0} by considering relevant classes of \emph{non-Poissonian} distributions of collision times and show how the Poisson statistics is recovered as a particular case. Once the master equation is constructed, we study the exact time evolution and the eventual appearance of the QZE over long time scales.

Details on the construction of the master equation corresponding to a general distribution of collision times are given in Section \ref{2}. Section \ref{3} is devoted to the the asymptotic dynamics of the populations of the energy levels and coherence for a general distribution of collision times and relevant particular cases. Time scales for inverse power laws are analytically estimated and the appearance of the QZE over long time scales is discussed. In Appendices \ref{A}, \ref{SAT} and \ref{B} a detailed analysis of the convoluted structure equations driving the dynamics is performed.

\section{The master equation for a general distributions of collision times}\label{2}

The model analyzed describes a multilevel molecule colliding in a gas of identical molecules. Each molecule is characterized by $N_L$ and $N_R$ internal energy levels of different parity, the subscripts $L$ and $R$ refer to "left" and "right", respectively. The ground levels are energy degenerate and coupled by a constant interaction. The collisions are assumed to conserve the spin parity, causing transitions between adjacent energy levels. The transitions between $L$ and $R$ spin states are forbidden, except through the ground states. The dynamics of the spatial degrees of freedom is not considered.

The total Hilbert space is spanned by the set of orthonormal state kets $\left\{\left|n_L\right.\rangle,\left|n_R\right.\rangle, \forall \,\,n_L=1,\ldots,N_L,\forall \,\, n_R=1,\ldots N_R \right\}$ where the kets $\left|n_L\right.\rangle$ and $\left|n_R\right.\rangle$ are eigenstates of left and right rotational energy levels spanning the left and right Hilbert subspaces, $\mathcal{H}_L$ and $\mathcal{H}_R$, respectively \cite{Zeno0}.
The total Hamiltonian is defined as follows:
\begin{eqnarray}
&&H=H_0+H_1, \label{Lf}\\
&&H_0=\sum_{n_L=1}^{N_L}E_{n_L}|n_L\rangle\langle
n_L|+\sum_{n_R=1}^{N_R}E_{n_R}|n_R\rangle\langle n_R|, \nonumber \\ &&H_1=\hbar \Omega \left(|1_L\rangle \langle 1_R| + |1_L\rangle
\langle 1_R|\right), \nonumber\\
&&  E_{1_L}=E_{1_R}=E_1, \hspace{1em}\langle1_L||1_R \rangle=0. \nonumber
 \end{eqnarray}
The term $H_1$ mimics the constant interaction between the ground states of different parity. The model refers to rotational energy levels, $E_{n_s}= \hbar \omega_s n_s\left(n_s+1\right)$, for every $n_s=1,2,\ldots N_s$ and $s=L,R$, where the ground is the only degenerate energy level, $E_{n_L}\neq E_{n^{\prime}_R}$ for every $n_L=2,\ldots,N_L$ and every $n_R^{\prime}=2,\ldots,N_R$.

The time evolution of the statistical density matrix describing the rotational states of the molecule in the Hilbert space of the total spin, corresponding to a Poisson distribution of collision times of mean $\tau_0$, is driven by the following Schr\"odinger equation in It\^o form:
\begin{equation}
\dot{ \rho}(t)= \left(\mathcal{L}_f
+\frac{1}{\tau_0}\mathcal{L}_c\right) \left[\rho(t)\right].
\label{MastEqPoisson}
\end{equation}
 The Liouville superoperator $L_f$ drives the free evolution, $\mathcal{L}_f\left[\cdot\right]=-\imath/\hbar\left[H,\cdot\right]$.   while the superoperator $\mathcal{L}_c$, related to the action of each collision, reads as follows:
\begin{eqnarray}
&&\mathcal{L}_c\left[\cdot\right]=-\imath\left[V,\cdot\right]
-\frac{1}{2}\left[V\left[V,\cdot\right]\right],\label{Lc}\\
&&V=\sum_{s=L,R} \alpha
_s\sum_{n_s=1}^{N_s-1} \left(|n_s\rangle\langle
n_{s+1}|+|n_s+1\rangle\langle n_s|\right). \nonumber
\end{eqnarray}
The structure of the interaction Hamiltonian $V$ recovers the assumption that the collisions provoke transitions between "nearest neighbor" energy levels of the same parity.

  The model multilevel molecule is now considered as an open quantum system and the collisions with other molecules, represent the interactions with the external environment and are described by the action of a superoperators on the density matrix of the molecule itself. This picture will be adopted for the construction of the master equation mimicking the dynamics of the internal energy levels of the model molecule in case of a general distribution of the collision times. By following the procedure of the CTQRW, a convoluted structure master equation is obtained for the time evolution of the internal energy levels for non-Poissonian distributions of collision times.
Recent applications of this technique are shown, for example, in Ref. \cite{BBGG} in order to interpret the fluorescence fluctuations in blinking quantum dots.

Following Refs. \cite{CTQRW0} and \cite{CTQRW1}, the CTQRW approach provides the following time evolution of the density matrix:
\begin{equation}
\rho(t)=\sum_{n=0}^{\infty}\int_0^t
\mathcal{P}_0\left(t-t^{\prime}\right)\Big[ W_n \left(t^{\prime}\right)
\left[\rho\left(0\right)\right]\Big] dt^{\prime},
 \label{CTQRWrho}
\end{equation}
   where the function $W_n(t)$ recalls the probability that $n$ interactions (collisions) has occurred, the last one at time $t$, while the function $\mathcal{P}_0(t)$ recalls the probability that no interaction occurs in time interval $t$.
For our model, $\mathcal{P}_0(t)$ and $W_n(t)$ are superoperators properly defined through their Laplace transforms \cite{CTQRW1}:
\begin{equation}
\tilde{\mathcal{P}}_0(u)=\tilde{P}_0\left(u-\mathcal{L}_0\right),
\hspace{1em} \tilde{W}_n(u)=\left(\Lambda
\tilde{w}\left(u-\mathcal{L}_0\right)\right)^n, \label{P0Wn}
\end{equation}
where the function $w(t)$ is the statistical distribution of the random intervals between two consecutive collisions.
  The superoperator $\Lambda$, mimicking the effect of each collision, is related  to the Liouvillian $\mathcal{L}_I$ by the relation: $\mathcal{L}_I=\Lambda-I$, where $I$ is the identity superoperator. In this way, the Laplace transform
of Eq. (\ref{CTQRWrho}) gives:
\begin{equation}
\tilde{\rho}(u)=\frac{I-\tilde{w}\left(u-\mathcal{L}_0\right)}
{u-\mathcal{L}_0}\left[
\frac{1}{I-\Lambda \tilde{w}\left(u-\mathcal{L}_0\right)}\left[ \rho(0)\right]\right],
\label{LaplaceGenMastEq}
\end{equation}
equivalent to the following non-Markovian master equation:
\begin{equation}
\dot{ \rho}(t)=\mathcal{L}_0 \left[\rho(t)\right]+\int_0^t
\Phi\left(t-t^{\prime}\right)\mathcal{L}_I\left[ e^{\mathcal{L}_0
\left(t-t^{\prime}\right)}
\left[\rho\left(t^{\prime}\right)\right]\right]dt^{\prime},
\label{GenMastEQ}
\end{equation}
where $\tilde{\Phi}=u \, \tilde{w}(u)/\left(1-\tilde{w}(u)\right)$ is the Laplace transform of the memory kernel related to the distribution of collision times. For more details we refer to \cite{CTQRW0,budiniPRE04,CTQRW1}.

In the Poisson case, the distribution of collision times and the corresponding memory kernel, respectively, read
\[
w_p(t)=\frac{e^{-t/\tau_0}}{\tau_0}, \hspace{2em} \Phi_p(t)=\frac{\delta(t)}{\tau_0}
 \]
 and the Gorini-Kossakowski-Sudarshan-Lindblad form of Eq. (\ref{MastEqPoisson}) is recovered from Eq. (\ref{GenMastEQ}).
The master equation (\ref{GenMastEQ}) describes the dynamics of the multilevel molecule undergoing the collision interactions (\ref{Lf}) with a generic distribution of the collision times, provided that $\mathcal{L}_0=\mathcal{L}_f$ and $\mathcal{L}_I=\mathcal{L}_c$. In this way, the following convoluted structure master equation is obtained:
\begin{equation}
\dot{ \rho}(t)=\mathcal{L}_f \rho(t)+\int_0^t
\Phi\left(t-t^{\prime}\right)\mathcal{L}_c \left[e^{\mathcal{L}_f
\left(t-t^{\prime}\right)}\left[\rho\left(t^{\prime}\right)\right]\right]dt^{\prime}.
\label{GenMastEq}
\end{equation}
The study of the above equation requires a preliminary analysis of the values of the parameters involved.
Let $\Delta E$ be the smallest energy difference between every couple of $L$ and $R$ states, the ground states being excluded. The estimates $\Delta E/\hbar\simeq 10^{-9} s$, $\Omega\simeq 1 kHz$ and $\tau_0\simeq 1 \mu s$ justify the assumption that all the level pairs are far from resonance condition,
\begin{equation}
\frac{\Delta E}{\hbar} \gg \max \left\{\Omega, \, \frac{1}{ \tau_{\Phi}}\right\},
\label{NH}
\end{equation}
  for every characteristic time $\tau_{\Phi}$ of the analyzed distributions of collision times, $\Phi(t)$ being the corresponding memory kernel.
 As pointed out in Ref. \cite{Zeno0}, the inequality (\ref{NH}) reveals the existence of a "fast" dynamics  giving a vanishing average contribution over long time scales. For a Poisson distribution of collision times, the dynamics of the populations of the energy levels plus the term mimicking the coherence between the ground states, is decoupled from the dynamics of the remaining coherence terms, evolving over different time scales. The same behavior appears for a general distribution of collision times, described by the equations built up in Appendix \ref{A} and recovered in Appendix \ref{SAT} through an adiabatic theorem \cite{AT}.

  Usually, the dissociation energy of the molecule is reached over time scales $T_d \simeq N^2 \tau_0/\left( \pi^2 \alpha^2\right)$, where the numbers $N$ and $\alpha$ are of the order of $N_s$ and $\alpha_s$, respectively, being $s=L,R$. Such time scale is much larger than the one characterizing the dynamics under study, for this reason the number of energy levels is finite, resulting, typically, in few tens.
   Thus, by considering $N_s\gg 10$, where $s=L,R$, the dynamics, described by a set of recursion equations, $N_L$ for the L subspace and $N_R$ for the R subspace, can be reasonably approximated by the solutions of two second order difference equations in the Laplace space, obtained by considering as infinite the number of both the energy levels $N_L$ and $N_R$. Details are given in Appendix \ref{B}. Notice also that the treatment is independent of the expressions of the energy levels $E_{n_L}$ and $E_{n_R}$.

   Roughly speaking, since the number of populated energy levels is confined to few tens, over the analyzed time scales, if $N_s\gg 10$, the dynamics is approximated by the one obtained for an infinite number of energy levels. A detailed analysis of the above estimates is given in Refs. \cite{Zeno0}, \cite{MSacp1958} and \cite{CM1980}.

\section{Non-Poissonian statistics and inverse power laws}\label{3}

This Section is devoted to the dynamics of the populations of the energy levels and to decoherence process.
We aim to describe the dynamics for relevant classes of non-Poisson distribution densities of collision times, recovering the Poisson statistic as a particular case.

The coherence term,
\[p^{c}(t)=\imath \left(\rho_{1_L 1_R}(t)-\rho_{1_R 1_L}(t)\right),\]
and the populations of the whole $L$ and $R$ levels,
\[P_s(t)= \sum_{n_s=1}^{N_s} p_{n_s}(t),\hspace{1em} s=L,R,\]
  are evaluated in Appendix \ref{A} for a generic distribution of collision times, through Eqs. (\ref{SimplGenMastEq1L1L}), (\ref{SimplGenMastEq1L1R}), (\ref{SimplGenMastEq2L2LD0}), (\ref{SimplGenMastEqmsmsD0}), (\ref{SimplGenMastEqNsNsD0}) and those obtained by exchanging the indexes $L$ and $R$, holding true for times $t\gg  1/\Omega$.

The detailed study performed in Appendix \ref{B}, reveals that the populations of the whole $L$ and $R$ energy levels are driven uniquely by the coherence term $p^c(t)$ and the populations of the ground levels:
\begin{eqnarray}
&&\hspace{-6em}\dot{P}_L(t)=\Omega p^c(t),\hspace{2em}\dot{P}_R(t)=-\Omega
p^c(t), \label{FinalEqs1}\\&&\hspace{-6em}\dot{p^c}(t)=2 \Omega
\left(p_{1_R}(t)-p_{1_L}(t)\right). \label{FinalEqs2}
\end{eqnarray}
The final value theorem \cite{widder}, applied to Eqs. (\ref{FinalEqs1}), (\ref{FinalEqs2}), (\ref{pc}) and (\ref{P1Lu}), suggests the existence of stable asymptotic configuration for \emph{every} distribution of collision times,
\begin{equation}
P_{L_{\left(R\right)}}\left(+\infty\right)=
\frac{\alpha_{L_{\left(R\right)}}}{\alpha_{L}+\alpha_{R}},
\hspace{2em}
\frac{P_{L}\left(+\infty\right)}{P_R\left(+\infty\right)}=
\frac{\alpha_L}{\alpha_R}. \label{PLdivPRasympt}
\end{equation}

The whole population of both the $L$ and $R$ levels tends to a stationary distribution depending uniquely on the collision interaction, characterized by the parameters $\alpha_L$ and $\alpha_R$, independent of either the Rabi-like oscillations between the ground levels or the statistical distribution of the random collision times. In case $\alpha_L=\alpha_R$, the asymptotic populations of the whole $L$ and $R$ energy levels become identical.

We now show in detail the exact dynamics of the levels populations and the coherence terms over long time scales, for relevant non-Poissonian distributions of collision times.

\subsection{Fractional diffusion}

As the first case, we consider the fractional diffusion processes widely studied in literature \cite{CTQRW0,S,Mainardi,MK}. The corresponding distribution density of collision times,
 \begin{equation}
w_r(t)=a_r^2 t^{-2 r} E_{1-2r,1-2r}\left(-a_r^2\,
t^{1-2r}\right),\hspace{1em}\frac{1}{2}>r > 0,\label{fractW}
\end{equation}
is defined through the generalized Mittag-Leffler function \cite{prabakar,MainardiBook}
\[
E_{\alpha,\beta}\left(z\right)=\sum_{n=0}^{\infty}
\frac{z^n}{\Gamma\left(\alpha n+\beta\right)},\hspace{2em}
z, \beta\in \mathbb{C}, \hspace{2em} \Re\left\{\alpha\right\}>0.
\]The mean time results to be infinite and the Poisson statistic is recovered for $r=0$.

The internal dynamics is analyzed through the Laplace transform of the corresponding memory kernel,
\begin{equation}
\tilde{\Phi}_r(u)=\frac{u \,\tilde{w}_r(u)}{1-\tilde{w}_r(u)}=a_r^2
u^{2r},\hspace{2em}\frac{1}{2}>r > 0,\label{fractWPhi}
\end{equation}
and the expression (\ref{pc}).
The following inverse power law behavior
\begin{equation}
p^{c}(t)\sim \frac{\left(\alpha_R-\alpha_L\right)\,t^{r-3/2}}{2\,
\Omega\, a_r \left(\alpha_L+\alpha_R\right)^2 \Gamma\left(r-1/2\right)},
\hspace{2em} t\to+\infty. \label{rhocasymptr}
\end{equation}
describes the dynamics over long time scales, $t \gg \tau_r$, defined below.
Initially, only the ground $L$ level is populated,
which means $P_L(0)=1$. For
$t \gg \tau_r$, Eqs. (\ref{FinalEqs1})
and (\ref{pc}) lead to the following dynamics of the whole population levels:
\begin{eqnarray}
&&P_{L_{\left(R\right)}}(t)\sim
\frac{\alpha_{L_{\left(R\right)}}}{\alpha_L+\alpha_R}
+_{(-)}
\frac{\left(\alpha_R-\alpha_L\right)\,t^{r-1/2}}{2
a_r \left(\alpha_L+\alpha_R\right)^2 \Gamma\left(r+1/2\right)},
\hspace{1em}t\to+\infty.
 \label{PLR0asympt}
\end{eqnarray}
The choice of the long time scale, in \emph{dimensionless} units, rises
from the convergence criteria concerning
the series expansion of Eq. (\ref{pc}),
\begin{eqnarray}
&&\tau_r=\max\Bigg\{\left(4 a_r^2\alpha_L^2\right)^{1/\left(2 r-1\right)},\left(4 a_r^2\alpha_R^2\right)^{1/\left(2 r-1\right)},
\left(\frac{\alpha_L^2+\alpha_R^2+3 \alpha_L\alpha_R}{a_r\alpha_L\alpha_R\left(\alpha_L+\alpha_R\right)}\right)^{2/\left(1-2r\right)}
\Bigg\}. \label{taur}
\end{eqnarray}
 For large collision time intervals, $t\gg a_r^{2/(2r-1)}$,
 the corresponding statistical distribution density is described by the asymptotic form
\begin{equation}
w_r(t)\sim \frac{(1-2r)t^{-2(1-r)}}{a_r^2
\,\Gamma\left(2 r\right)}, \hspace{2em} t\rightarrow+\infty.
\nonumber
\end{equation}

An increase of the power law tail of the distribution density (\ref{fractW}) arbitrary slows down the relaxations, in the limit $r\to \frac{1}{2}^{-}$. Hence, an increase of consecutive collisions over long time intervals hinders the long time dynamics. This behavior may be interpreted as the appearance of the QZE over \emph{long} time scales.

\subsection{Inverse Power law distribution}

 We now consider a power law distribution density of collision times,
 \begin{equation}
w_{\mu}(t)=\frac{\left(\mu-1\right)T^{\mu-1}}
{\left(t+T\right)^{\mu}}. \hspace{2em}
2>\mu>1, \label{powerlaw}
\end{equation}
 The mean time results to be infinite. This case has been widely studied in
literature in both classical \cite{ZK} and quantum
processes \cite{BBGG}. The behavior of the Laplace transform of the distribution of collision times in the origin of the complex plane,
\begin{equation}
\tilde{w}_{\mu}(u)
\sim 1-\Gamma\left(2-\mu\right)
\left(uT\right)^{\mu-1},
\hspace{2em}
u \rightarrow 0^+, \label{pl1}
\end{equation}
gives the following relationship about the memory kernel:
\begin{equation}
 \tilde{\Phi}_{\mu}(u)=\frac{u \,\tilde{w}_{\mu}(u) }{1-\tilde{w}_{\mu}(u)}\sim
\frac{u^{2 -\mu}}{\Gamma\left(2-\mu\right) T^{\mu-1}},\hspace{2em}
u \rightarrow 0^+. \label{pl2}
\end{equation}
   The analysis of the power series expansion of Eq. (\ref{pc}) in $u=0$ and the sector of convergence in the complex plane leads to
   inverse power law behaviors over a long time scale, $t\gg \tau_{\mu}$, defined below,
\begin{eqnarray}
&&\tilde{p}^c(t)\sim \frac{\left(\alpha_R-\alpha_L\right)
\sqrt{\Gamma\left(2-\mu\right)}}{2 \, \Omega \, T \left(\alpha_L+\alpha_R\right)^2\Gamma\left(\left(1-\mu\right)/2\right)}\nonumber \\ &&\times \left(\frac{t}{T}\right)^{-\left(1+\mu\right)/2}, \hspace{2em} t \to+\infty, \label{pcasymptmu}
\\
&&P_{L_{\left(R\right)}}(t)\sim
\frac{\alpha_{L_{\left(R\right)}}}{\alpha_L+\alpha_R}+_{(-)}
\frac{\left(\alpha_R-\alpha_L\right)
\sqrt{\Gamma\left(2-\mu\right)}}{2 \left(\alpha_L+\alpha_R\right)^2
\Gamma\left(\left(3-\mu\right)/2\right)}
\nonumber \\&&\times \left(\frac{t}{T}\right)^{\left(1-\mu\right)/2},\hspace{2em}t\to +\infty.\label{PLRasymptPowerlaw}
\end{eqnarray}
 The
 convergence criteria for the series expansion of Eq. (\ref{pc}), lead to the following definition of long time scale in \emph{dimensionless} units:
\begin{eqnarray}
&&\tau_{\mu}=\max\Bigg\{T,T \left(\frac{\alpha_L^2+\alpha_R^2+3 \alpha_L\alpha_R}{\left(\Gamma\left(2-\mu\right)\right)^{1/2}
\alpha_L\alpha_R\left(\alpha_L+\alpha_R\right)}\right)^{2/\left(\mu-1\right)},
\nonumber \\&&
T \left(\frac{\Gamma\left(2-\mu\right)\left(1+4 \alpha_L^2\right)}{8 \alpha_L^2}\right)^{1/\left(\mu-1\right)} ,T \left(\frac{\Gamma\left(2-\mu\right)\left(1+4 \alpha_R^2\right)}{8 \alpha_R^2}\right)^{1/\left(\mu-1\right)}
\Bigg\}. \,\,\,\,\,\,\,\,\,\,\,\,\,\,\,\label{taumu}
\end{eqnarray}

A decrease of the power $\mu$ provokes an increase of the distribution density of the random collisions over sufficiently long time intervals. Consequently, the long time relaxations become arbitrarily slow in the limit $\mu\to1^+$ and the long time dynamics is hindered.
Again, this behavior may be interpreted as the appearance of the QZE over \emph{long} time scales.

\subsection{Exponential memory kernel}

The case of exponential kernel, $\Phi_{\gamma}(t)=A e^{- \gamma t} $, is an
argument of interest in stochastic processes \cite{CTQRW0}. The corresponding distribution of collision times is
\begin{equation}
w_{\gamma}(t)=2 A  \frac{\sinh
\left(t\sqrt{\gamma^2-4 A}\right)e^{-\gamma t/2}}{\sqrt{\gamma^2-4A}},\hspace{1em}
\gamma^2>4A,
\label{wexpkernl}
\end{equation}
with finite mean time $T_{\gamma}=\gamma/A$.

The asymptotic dynamics of the levels populations and the coherence term are obtained from the Laplace transform of the memory kernel, $\tilde{\Phi}_{\gamma}(u)=
A/\left(\gamma+u\right)$,
and the analysis Eq. (\ref{pc}).
With this method, the following asymptotic forms are obtained:
\begin{eqnarray}
&&\tilde{p}^c(t)\sim\frac{\left(\alpha_R-\alpha_L\right)\, }{2 \,\Omega\, T_{\gamma}\left(\alpha_L+\alpha_R\right)^2 \Gamma\left(-1/2\right)}\nonumber \\ &&\times\left(\frac{t}{T_{\gamma}}\right)^{-3/2}, \hspace{2em} t\to+\infty,\label{pcasymptgamma}
\\
&&P_{L_{\left(R\right)}}(t)\sim \frac{\alpha_{L_{\left(R\right)}}}
{\alpha_L+\alpha_R}+_{\left(-\right)}
\frac{\left(\alpha_L-\alpha_R\right)
}{ 2 \left(\alpha_L+\alpha_R\right)^2
\Gamma\left(1/2\right)} \nonumber \\
&&\times\left(\frac{t}{T_{\gamma}}\right)^{-1/2}, \hspace{2em} t\to+\infty,
\label{ExpCollPopul}
\end{eqnarray}
describing the time evolutions of the coherence term and the whole $L$ and $R$ levels populations, respectively, over a long time scale, $t \gg \tau_{\gamma}$, defined below.

 The convergence criteria of the series expansion of Eq. (\ref{pc}), lead to the following choice of the long time scale defined in \emph{dimensionless} units:
\begin{eqnarray}
&&\tau_{\gamma}=\max
\Bigg\{\frac{1}{\gamma}+\frac{T_{\gamma}}{4 \alpha_L^2},\frac{1}{\gamma}+\frac{T_{\gamma}}{4 \alpha_R^2},T_{\gamma}\left(\frac{\alpha_L^2+\alpha_R^2+3 \alpha_L\alpha_R}{\alpha_L\alpha_R\left(\alpha_L+\alpha_R\right)}\right)^2\Bigg\}.\label{taugamma}
\end{eqnarray}

\subsection{Bi-exponential distribution}

We now consider the case of a bi-exponential \cite{CTQRW1}
distribution of collision times
\begin{eqnarray}
&&
w_{be}(t)=P_a D_a e^{-D_a t}+P_b D_b e^{-D_b t}, \label{MTbiexp} \\ &&P_a,D_a,P_b,D_b>0,\hspace{1em}
P_a+P_b=1, \nonumber
\end{eqnarray}
with finite mean time
\[
T_{be}=\frac{P_a}{D_a}+\frac{P_b}{D_b}.
\]
As in the previous cases, we analyze the expression (\ref{pc}) corresponding to the memory kernel
\[
 \tilde{\Phi}_{be}(u)=\frac{D_a D_b+u \left(D_a P_a
+D_b P_b\right)} {D_a P_b +D_b P_a+u}.
\]
The following asymptotic forms emerge:
\begin{eqnarray}
&&p^{(c)}(t)\sim\frac{\left(\alpha_R-
\alpha_L\right)}{ 2 \,\Omega \,T_{be}
\left(\alpha_L+\alpha_R\right)^2
\Gamma\left(-1/2\right)} \nonumber \\
&&\times \left(\frac{t}{T_{be}}\right)^{-3/2}, \hspace{2em} t\to+\infty, \label{pcasymptbe}\\
&&P_{L_{\left(R\right)}}(t)\sim \frac{\alpha_{L_{\left(R\right)}}}
{\alpha_L+\alpha_R}+_{\left(-\right)}
\frac{\left(\alpha_R-\alpha_L\right) }{
\left(\alpha_L+\alpha_R\right)^2
\Gamma\left(1/2\right)} \nonumber \\
&&\times \left(\frac{t}{T_{be}}\right)^{-1/2}, \hspace{2em}
t\to+\infty,\label{BiExpCDAsymptPopul}
\end{eqnarray}
describing the dynamics over long a time scale, $t \gg \tau_{be}$, defined below.
The convergence criteria of the series expansion of Eq. (\ref{pc}) lead to the following time scale defined in \emph{dimensionless} units:
\begin{eqnarray}
&&\tau_{be}=\max\Bigg\{  q+ \frac{T_{be}}{4 \alpha_L^2}+\frac{1}{d},q+ \frac{T_{be}}{4 \alpha_R^2}+\frac{1}{d},\nonumber \\ &&
\hspace{5.5em}T_{be}
\left(\frac{\alpha_L^2+\alpha_R^2+3 \alpha_L\alpha_R}{\alpha_L\alpha_R\left(\alpha_L+\alpha_R\right)}\right)^2\Bigg\}, \label{taube}
\end{eqnarray}
where
\[q=\frac{P_a}{D_b}+\frac{P_b}{D_a},\hspace{1em}
d=D_a P_b+D_b P_a.
\]

 Notice that the particular case $P_a=1$ and $P_b=0$ gives the Poisson distribution of collision times$, w_p(t)=e^{- t/ \tau_0}/\tau_0$,  and it is described by the master equation (\ref{MastEqPoisson}), where $T_{be}=1/D_a$.

\section{Concluding remarks}

   A gas of identical multilevel molecules is modeled by two finite sets of rotational internal energy levels of different parity and degenerate ground states, coupled by a constant interaction. The prescriptions of the CTQRW is adopted to build up the master equation driving the exact dynamics of the rotational internal energy levels for a general distribution of collision times. The resulting master equation turns out to be of convoluted structure and recovers the Gorini-Kossakowski-Sudarshan-Lindblad form for the Poisson statistics.

The populated energy levels of the molecule are typically few tens and far from the dissociation energy, over the analyzed time scales. Thus, by considering $N_s\gg10$, for $s=L,R$, over such time scales, the dynamics can be reasonably approximated by the equations of motion obtained for an infinite number of energy levels. In this way, the resulting population of the whole $L$ and $R$ levels tends to a unique stable equilibrium configuration, for every random distribution of the collision times. The time evolution of both the populations and the decoherence term exhibits inverse power law behavior over estimated long time scales, for bi-exponential and power law distributions of collision times, for fractional diffusion and exponential memory kernel.

The cases with infinite mean collision times provide inverse power law relaxations that may become arbitrarily slow. We observe that the long time dynamics is hindered by increasing the distribution density of the random collisions over long time intervals, as the tail approaches the power law $1/t$. This behavior may be interpreted as a QZE over \emph{long} time scales.

\appendix
\section{The general master equation in details}\label{A}

 This Appendix is dedicated to the analysis of the dynamics
 described by Eq. (\ref{GenMastEq}). Since the ground states are degenerate, the Hamiltonian $H$ takes the following diagonal form:
\begin{eqnarray}
&& \hspace{-7em}H=E_{1_-} |1_-\rangle\langle 1_-|  + E_{1_+}
|1_+\rangle\langle1_+| +\sum_{s=L,R}\sum_{n_s=2}^{N_s} E_{n_s}
|n_s\rangle\langle n_s|,\label{diagonalH}  \\
&& \hspace{-7em}|1_{L_{\left(R\right)}}\rangle=\gamma^{\left(+\right)}_{L_{\left(R\right)}}
|1_+ \rangle + \gamma^{\left(-\right)}_{L_{\left(R\right)}}|1_-\rangle,\hspace{2em}\langle 1_+|1_-\rangle=0, \nonumber \\
&& \hspace{-7em}|1_{\pm}\rangle=g^{\left(\pm\right)}_{R}|1_R\rangle +
g^{\left(\pm\right)}_{L}|1_L\rangle,
\hspace{2em}E_{1}\pm =E_1 \pm \hbar \Omega,\nonumber\\&&
   \hspace{-7em}g^{\left(\pm\right)}_R=\gamma^{\left(\pm\right)}_R=\frac{ \pm1}{\sqrt{2}
}, \hspace{2em}
g^{\left(\pm\right)}_L=\gamma^{\left(\pm\right)}_L=\frac{ 1}{\sqrt{2}
}. \nonumber
\end{eqnarray}
After long but straightforward algebra the
time evolution of each element of the statistical density matrix is obtained. The frequency terms are labeled as $\omega_{\nu, \eta}=\left(E_{\nu}-E_{\eta}\right)/ \hbar$.
We start from the evaluation of the master equation driving the dynamics of matrix element $\rho_{1_L 2_L}(t)$,
\begin{eqnarray}
&&\hspace{-4em}\dot{ \rho}_{1_L, 2_L}(t)=\imath \left(\omega_{2_L 1_L} \rho_{1_L
2_L}(t)- \Omega \rho_{1_R
2_L}(t)\right)+ \sum_{i=L,R}\sum_{j=L,R}\sum_{n=1}^4 \sum_{m=1}^4 \nonumber \\&&\hspace{1.2em}\int_0^t
\Phi\left(t-t^{\prime}\right)
\varphi^{\left(1_L,2_L\right)}_{n_i,m_j}\left(t-t^{\prime}\right)
\rho_{n_i,m_j}\left(t^{\prime}\right) dt^{\prime},
\label{GenMastEq1L2L}
\end{eqnarray}
where the non-vanishing terms
$\varphi^{\left(1_L,2_L\right)}_{n_i,m_j}\left(\tau\right)$ are
\begin{eqnarray}
&&\varphi^{\left(1_L,2_L\right)}_{1_L,1_L}\left(\tau\right)= \imath
\alpha_L
\cos^2\left(\Omega\tau\right),\hspace{0.5em}
\varphi^{\left(1_L,2_L\right)}_{1_R,1_R}\left(\tau\right)=
\imath \alpha_L \sin^2\left(\Omega\tau\right),\hspace{0.5em}
\varphi^{\left(1_L,2_L\right)}_{1_L,1_R}\left(\tau\right)\nonumber \\ &&=
-\left(\varphi^{\left(1_L,2_L\right)}_{1_R,1_L}\left(\tau\right)
\right)^{\ast}=
-  \frac{\alpha_L}{2}\sin\left(2 \Omega\tau\right),\hspace{0.5em}
\varphi^{\left(1_L,2_L\right)}_{1_L,2_L}\left(\tau\right)= -\frac{3}{2}
\,\alpha_L^2 e^{\imath \omega_{2_L 1_L}\tau} \nonumber \\&& \times\,
\cos\left(\Omega\tau\right) , \hspace{0.5em}
\varphi^{\left(1_L,2_L\right)}_{1_L,3_L}\left(\tau\right)=
 \imath \alpha_L e^{\imath \omega_{3_L 1_L}\tau}
 \cos\left(\Omega\tau\right), \hspace{0.5em}
\varphi^{\left(1_L,2_L\right)}_{1_L,4_L}\left(\tau\right)\nonumber \\&&=
-\frac{\alpha_L^2}{2} \,e^{\imath \omega_{4_L 1_L} \tau }
\cos\left(\Omega \tau\right),\hspace{0.5em}
\varphi^{\left(1_L,2_L\right)}_{3_L,2_L}\left(\tau\right)=
-\frac{\alpha_L^2}{2}
\,e^{-\imath \omega_{3_L 2_L} \tau},
\nonumber \\ &&\varphi^{\left(1_L,2_L\right)}_{2_L,1_L}\left(\tau\right)=-\frac{2}{3}
\left(\varphi^{\left(1_L,2_L\right)}_{1_L,2_L}
\left(\tau\right)\right)^{\ast},\hspace{0.5em}
\varphi^{\left(1_L,2_L\right)}_{1_R,2_L}\left(\tau\right)= \frac{3}{2}\,\imath\,
 \alpha_L^2\sin\left(\Omega \tau\right)\nonumber \\&&\times \,e^{\imath \omega_{2_L 1_L}\tau}=-\frac{2}{3}
\left(\varphi^{\left(1_L,2_L\right)}_{2_L,1_R}
\left(\tau\right)\right)^{\ast},\hspace{0.5em}
\varphi^{\left(1_L,2_L\right)}_{1_R,3_L}\left(\tau\right)= \alpha_L \sin\left(\Omega\tau\right)\nonumber \\&&\times \,e^{\imath \omega_{3_L 1_L}\tau},\hspace{0.5em}
\varphi^{\left(1_L,2_L\right)}_{2_L,2_L}\left(\tau\right)=-\imath
\alpha_L,\hspace{1em}
\varphi^{\left(1_L,2_L\right)}_{1_R,4_L}\left(\tau\right)=
-\frac{\alpha_L^2}{4} e^{\imath\left(\omega_{4_L 1_L}-\Omega\right) \tau},\nonumber \\ &&\varphi^{\left(1_L,2_L\right)}_{2_L,3_L}
\left(\tau\right)= e^{\imath
\omega_{3_L 2_L}\tau} \alpha_L^2, \hspace{1em}\varphi^{\left(1_L,2_L\right)}_{2_L,1_L}\left(\tau\right)= -
\alpha_L^2 e^{-\imath \omega_{2_L 1_L}\tau}\cos\left(\Omega \tau\right) \nonumber.
\end{eqnarray}

The term $\dot{\rho}_{1_L,1_L}(t)$ reads
\begin{eqnarray}
&&\hspace{-4em}\dot{\rho}_{1_L,1_L}(t)=\imath \Omega \left(\rho_{1_L 1_R}(t)-
\rho_{1_R 1_L}(t)\right)+\sum_{i=L,R}\sum_{j=L,R}\sum_{n=1}^3
\sum_{m=1}^3 \nonumber \\ &&\hspace{1em}\int_0^t \Phi\left(t-t^{\prime}\right)
\varphi^{\left(1_L,1_L\right)}_{n_i,m_j}\left(t-t^{\prime}\right)
\rho_{n_i,m_j}\left(t^{\prime}\right) dt^{\prime},
\label{GenMastEq1L1L}
\end{eqnarray}
and the non-vanishing terms
$\varphi^{\left(1_L,1_L\right)}_{n_i,m_j}\left(\tau\right)$ are listed below:
\begin{eqnarray}
&&\varphi^{\left(1_L,1_L\right)}_{1_L,1_L}\left(\tau\right)=
-\alpha_L^2
\cos^2\left(\Omega\tau\right),\hspace{0.5em}
\varphi^{\left(1_L,1_L\right)}_{1_R,1_R}\left(\tau\right)=
-\alpha_L^2
\sin^2\left(\Omega\tau\right),\nonumber \\ &&\varphi^{\left(1_L,1_L\right)}_{1_L,1_R}\left(\tau\right)=
-\imath \frac{\alpha_L^2}{2}
\sin\left(2\Omega\tau\right), \hspace{0.5em}
\varphi^{\left(1_L,1_L\right)}_{2_L,2_L}\left(\tau\right)=\alpha_L^2, \nonumber \\ &&
\varphi^{\left(1_L,1_L\right)}_{1_L,3_L}\left(\tau\right)=
-\frac{\alpha_L^2}{2}\,e^{\imath\omega_{3_L 1_L} \tau} \cos\left(\Omega\tau\right)=
\left(\varphi^{\left(1_L,1_L\right)}_{3_L,1_L}
\left(\tau\right)\right)^{\ast}
, \hspace{0.5em}\varphi^{\left(1_L,1_L\right)}_{1_L,2_L}\left(\tau\right)\nonumber \\&&=\left(\varphi^{\left(1_L,1_L\right)}_{2_L,1_L}
\left(\tau\right)\right)^{\ast}= \imath \alpha_L
e^{\imath \omega_{2_L 1_L}\tau} \cos\left(\Omega\tau\right),\hspace{0.5em}
\varphi^{\left(1_L,1_L\right)}_{1_R,2_L}\left(\tau\right)=\alpha_L \sin\left(\Omega\tau\right)\nonumber \\ &&\times\, e^{\imath \omega_{2_L
1_L}\tau}=\left(\varphi^{\left(1_L,1_L\right)}_{2_L,1_R}
\left(\tau\right)\right)^{\ast}, \hspace{0.5em}
\varphi^{\left(1_L,1_L\right)}_{1_R,3_L}\left(\tau\right)=\left(\varphi^{\left(1_L,1_L\right)}_{3_L,1_R}
\left(\tau\right)\right)^{\ast}\nonumber \\ &&=
 \imath\frac{  \alpha^2_L }{2}\,\sin\left(\Omega\tau\right)e^{\imath\omega_{3_L
1_L}\tau}.
 \nonumber
  \end{eqnarray}
The term $\dot{\rho}_{1_R, 1_R}(t)$ is obtained by exchanging the subscripts $R$ and $L$.

As regards the coherent term $\dot{\rho}_{1_L 1_R}(t)$, we obtain
\begin{eqnarray}
&&\hspace{-5em}\dot{\rho}_{1_L, 1_R}(t)=\imath \Omega \left(\rho_{1_L 1_L}(t)-
\rho_{1_R 1_R}(t)\right)+\sum_{i=L,R}\sum_{j=L,R}\sum_{n=1}^3 \sum_{m=1}^3 \nonumber \\ &&\hspace{0.3em}\int_0^t
\Phi\left(t-t^{\prime}\right)
\varphi^{\left(1_L,1_R\right)}_{n_i,m_j}\left(t-t^{\prime}\right)
\rho_{n_i,m_j}\left(t^{\prime}\right) dt^{\prime},
\label{GenMastEq1L1L}
\end{eqnarray}
with the following non-vanishing terms:
\begin{eqnarray}
&&\varphi^{\left(1_L,1_R\right)}_{1_L,1_L}\left(\tau\right)=
 \frac{-\imath \left(\alpha_L^2+\alpha_R^2\right)\sin\left(2\Omega\tau\right)}{4}
 = -\varphi^{\left(1_L,1_R\right)}_{1_R,1_R}\left(\tau\right), \hspace{0.5em} \varphi^{\left(1_L,1_R\right)}_{1_L,1_R}\left(\tau\right)\nonumber \\ &&=
 -\frac{\alpha_L^2+\alpha_R^2}{2} \cos\left(2 \Omega
\tau\right), \hspace{0.5em}
\varphi^{\left(1_L,1_R\right)}_{1_R,1_L}
\left(\tau\right)= -\frac{
\left(\alpha_R^2+\alpha_L^2\right)}{2}  \,\sin^2\left(\Omega\tau\right), \nonumber
\\&&
\varphi^{\left(1_L,1_R\right)}_{1_R,2_R}\left(\tau\right)=\alpha_R \sin\left(\Omega\tau\right)e^{\imath\omega_{2_R
1_R}\tau},\hspace{1em}\varphi^{\left(1_L,1_R\right)}_{1_R,3_R}\left(\tau\right)=
\frac{\imath}{2}
\alpha_R^2 e^{\imath\omega_{3_R 1_R}\tau} \nonumber \\&&
\times\,\sin\left(\Omega\tau\right),\hspace{0.5em}
 \varphi^{\left(1_L,1_R\right)}_{1_L,2_R}
\left(\tau\right) =\imath\alpha_R
 e^{\imath\omega_{2_R 1_R}
\tau}\cos\left(\Omega\tau\right), \nonumber\\
 &&\varphi^{\left(1_L,1_R\right)}_{1_L,3_R}\left(\tau\right)=
-\frac{\alpha^2_R}{2}e^{\imath\omega_{3_R 1_R}\tau}
 \cos\left(\Omega\tau\right),\hspace{0.5em}
 \varphi^{\left(1_L,1_R\right)}_{2_L,1_L}\left(\tau\right)=
  \alpha_L e^{\imath\omega_{2_L
1_L}\tau} \nonumber \\ &&\times\, \sin\left(\Omega\tau\right),\hspace{0.5em}
\varphi^{\left(1_L,1_R\right)}_{2_L,2_R}\left(\tau\right)= \alpha_R
\alpha_L e^{\imath \left(\omega_{2_R 1_R}-\omega_{2_L
1_L}\right)\tau},\hspace{0.5em} \varphi^{\left(1_L,1_R\right)}_{2_L,1_R}
\left(\tau\right)\nonumber \\ &&=-\imath\alpha_L
 e^{-\imath\omega_{2_L 1_L}\tau}\cos\left(\Omega\tau\right), \hspace{0.5em}\varphi^{\left(1_L,1_R\right)}_{3_L,1_L}
 \left(\tau\right)= -\frac{\imath}{2}\,
\alpha_L^2  \sin\left(\Omega\tau\right) e^{-\imath\omega_{3_L
1_L}\tau},\nonumber
\\&&
\varphi^{\left(1_L,1_R\right)}_{3_L,1_R}\left(\tau\right)=
-\frac{\alpha^2_L}{2}\,e^{-\imath\omega_{3_L 1_L}\tau}
\cos\left(\Omega\tau\right)
. \nonumber
\end{eqnarray}

The term $\dot{\rho}_{2_L,2_L}(t)$ reads
\begin{equation}
\dot{ \rho}_{2_L, 2_L}(t)=\sum_{i=L,R}\sum_{j=L,R}\sum_{n=1}^4
\sum_{m=1}^4 \int_0^t \Phi\left(t-t^{\prime}\right)
\varphi^{\left(2_L,2_L\right)}_{n_i,m_j}\left(t-t^{\prime}\right)
\rho_{n_i m_j}\left(t^{\prime}\right) dt^{\prime},
\label{GenMastEq2L2LD0}
\end{equation}
with non-vanishing terms
\begin{eqnarray}
&&\varphi_{1_L,1_L}^{\left(2_L,2_L\right)}\left(\tau\right)=
\alpha_L^2\cos\left(\Omega \tau\right),\hspace{0.5em}
\varphi_{1_L,1_R}^{\left(2_L,2_L\right)}\left(\tau\right)= \imath
\frac{\alpha_L^2}{2}\sin\left(2\Omega
\tau\right)\nonumber
\\
&&=\left(\varphi_{1_R,1_L}^{\left(2_L,2_L\right)}
\left(\tau\right)\right)^{\ast},\hspace{0.5em}
\varphi_{1_L,2_L}^{\left(2_L,2_L\right)}\left(\tau\right)=- \imath
\alpha_L e^{\imath \omega_{2_L,1_L}\tau}\cos\left(\Omega
\tau\right)\nonumber \\&&=
\left(\varphi_{2_L,1_L}^{\left(2_L,2_L\right)}
\left(\tau\right)\right)^{\ast},\hspace{0.5em}
\varphi_{1_L,3_L}^{\left(2_L,2_L\right)}\left(\tau\right)=
\alpha_L^2 e^{\imath \omega_{3_L,1_L}\tau}\cos\left(\Omega
\tau\right),\hspace{0.5em}\varphi_{1_R,1_R}^{\left(2_L,2_L\right)}
\left(\tau\right)\nonumber\\ &&=
\alpha_L^2 \sin^2\left(\Omega \tau\right),\hspace{0.5em}
\varphi_{1_R,3_L}^{\left(2_L,2_L\right)}\left(\tau\right)=-
\imath\alpha_L^2e^{\imath \omega_{3_L,1_L}\tau}\sin\left(\Omega
\tau\right),\hspace{0.5em}
\varphi_{1_L,3_L}^{\left(2_L,2_L\right)}\left(\tau\right)\nonumber \\&&=
\alpha_L^2 e^{\imath \omega_{3_L,1_L}\tau}\cos\left(\Omega
\tau\right),\hspace{0.5em}\varphi_{1_R,2_L}^{\left(2_L,2_L\right)}
\left(\tau\right)=\varphi_{2_L,1_R}^{\left(2_L,2_L\right)}
\left(\tau\right)=-
 \alpha_L e^{-\imath \omega_{2_L,1_L}\tau}\nonumber \\ &&\times \,\sin\left(\Omega
\tau\right),\hspace{0.5em}\varphi_{2_L,2_L}^{\left(2_L,2_L\right)}
\left(\tau\right)=-3 \alpha_L^2, \hspace{0.5em}
\varphi_{2_L,3_L}^{\left(2_L,2_L\right)} \left(\tau\right)=\imath
\alpha_L e^{\imath \omega_{3_L,2_L}\tau},\nonumber
\\
&&
\varphi_{2_L,4_L}^{\left(2_L,2_L\right)} \left(\tau\right)=-
\alpha_L^2 e^{\imath \omega_{4_L,2_L}\tau},\hspace{0.5em}
\varphi_{3_L,1_L}^{\left(2_L,2_L\right)}\left(\tau\right)=\alpha_L^2
e^{-\imath \omega_{3_L,1_L}\tau} \cos\left(\Omega \tau\right),\nonumber \\ &&
 \varphi_{3_L,1_R}^{\left(2_L,2_L\right)}\left(\tau\right)=\imath
\alpha_L^2 e^{-\imath \omega_{3_L,1_L}\tau}\sin\left(\Omega
\tau\right),\hspace{0.5em}
\varphi_{3_L,2_L}^{\left(2_L,2_L\right)}\left(\tau\right)=-\imath
\alpha_L e^{-\imath
\omega_{3_L,2_L}\tau},\nonumber \\ &&\varphi_{4_L,2_L}^{\left(2_L,2_L\right)}
\left(\tau\right)=-\frac{\alpha_L^2}{2}\, e^{-\imath
\omega_{4_L,2_L}\tau}.\nonumber
\end{eqnarray}
The terms
$\dot{\rho}_{1_R, 1_R}(t)$, $\dot{\rho}_{1_R,1_L}(t)$,
$\dot{\rho}_{2_R, 2_R}(t)$,
$\dot{\rho}_{3_L,3_L}(t)$ are obtained by exchanging the indexes $L$ and $R$. We write down only the term $\dot{\rho}_{3_L, 3_L}(t)$ for the sake of shortness,
\begin{equation}
\dot{\rho}_{3_L, 3_L}(t)=\sum_{i=L,R}\sum_{j=L,R}\sum_{n=1}^4
\sum_{m=1}^5 \int_0^t \Phi\left(t-t^{\prime}\right)
\varphi^{\left(3_L,3_L\right)}_{n_i,m_j}\left(t-t^{\prime}\right)
\rho_{n_i m_j}\left(t^{\prime}\right) dt^{\prime},
\label{GenMastEq3L3LD0}
\end{equation}
and the corresponding non-vanishing terms,
\begin{eqnarray}
&&\varphi_{1_L,3_L}^{\left(3_L,3_L\right)}\left(\tau\right)=
\left(\varphi_{3_L,1_L}^{\left(3_L,3_L\right)}
\left(\tau\right)\right)^{\ast}=-\frac{ \alpha_L^2 }{2}\,  e^{\imath
\omega_{3_L,1_L}\tau}\cos\left(\Omega \tau\right),\hspace{0.5em}
\varphi_{1_R,3_L}^{\left(3_L,3_L\right)}\left(\tau\right)\nonumber \\ &&=\imath\frac{
\alpha_L^2 }{2}\,e^{\imath \omega_{3_L,1_L}\tau}  \sin\left(\Omega
\tau\right),\hspace{0.5em}
\varphi_{2_L,2_L}^{\left(3_L,3_L\right)}\left(\tau\right)
=\varphi_{4_L,4_L}^{\left(3_L,3_L\right)}\left(\tau\right)=\alpha_L^2,\hspace{0.5em}
\varphi_{2_L,3_L}^{\left(3_L,3_L\right)}\left(\tau\right)\nonumber \\&&=
\left(\varphi_{3_L,2_L}^{\left(3_L,3_L\right)}\left(\tau\right)
\right)^{\ast}=-\imath \alpha_L e^{\imath \omega_{3_L,2_L} \tau},
\hspace{0.5em}
\varphi_{3_L,5_L}^{\left(3_L,3_L\right)}\left(\tau\right)=
- \frac{\alpha_L^2}{2}\, e^{-\imath \omega_{5_L,3_L}
\tau},\nonumber
\\
&&\varphi_{2_L,4_L}^{\left(3_L,3_L\right)}\left(\tau\right)=
\left(\varphi_{4_L,2_L}^{\left(3_L,3_L\right)}
\left(\tau\right)\right)^{\ast}=\alpha_L^2
e^{\imath \omega_{4_L,2_L} \tau} ,\hspace{0.5em}
\varphi_{3_L,1_R}^{\left(3_L,3_L\right)}
\left(\tau\right)=-\imath\frac{
\alpha_L^2}{2} \nonumber \\&& e^{-\imath \omega_{3_L,1_L}\tau} \sin\left(\Omega
\tau\right), \hspace{0.5em}
\varphi_{3_L,3_L}^{\left(3_L,3_L\right)}
\left(\tau\right)=-2\alpha_L^2, \hspace{0.5em}\varphi_{3_L,4_L}^{\left(3_L,3_L\right)}\left(\tau\right)=
\left(\varphi_{4_L,3_L}^{\left(3_L,3_L\right)}\left(\tau\right)\right)^{\ast}\nonumber \\&&=
\imath \alpha_L e^{\imath \omega_{4_L,3_L} \tau}.
 \nonumber
 \end{eqnarray}

The terms $\dot{\rho}_{m_s, m_s}(t)$, are described by the following forms:
\begin{eqnarray}
&&\dot{\rho}_{m_s,
m_s}(t)=\sum_{i=m_s-2}^{m_s+2}\sum_{j=m_s-2}^{m_s+2}
\int_0^t\Phi\left(t-t^{\prime}\right)
\varphi^{\left(m_s,m_s\right)}_{i,j}\left(t-t^{\prime}\right)
\rho_{i,j}\left(t^{\prime}\right) d t^{\prime},\,\,\,\,\,\,\,\,\,\,\,\,\,\,\, \label{GenMastEqmsms}\\
&&4 \leq m_s \leq N_s-2,\hspace{1em}s=L,R, \nonumber
\end{eqnarray}
with non-vanishing terms
\begin{eqnarray}
&&\varphi^{\left(m_s,m_s\right)}_{m_s-2,m_s}\left(\tau\right)=
-\frac{\alpha_s^2}{2}\,
e^{\imath\left(E_{m_s}-E_{m_s-2}\right)\tau/\hbar},\hspace{0.5em}
\varphi^{\left(m_s,m_s\right)}_{m_s-1,m_s-1}\left(\tau\right)=
\alpha_s^2,\nonumber \\ &&
\varphi^{\left(m_s,m_s\right)}_{m_s-1,m_s}\left(\tau\right)= -\imath
\alpha_s
e^{\imath\left(E_{m_s}-E_{m_s-1}\right)\tau/\hbar},\hspace{0.5em}
\varphi^{\left(m_s,m_s\right)}_{m_s-1,m_s+1}\left(\tau\right)\nonumber \\ &&
=\alpha_s^2
e^{\imath\left(E_{m_s+1}-E_{m_s-1}\right)\tau/\hbar},\hspace{0.5em}
\varphi^{\left(m_s,m_s\right)}_{m_s,m_s-2}\left(\tau\right)=
-\frac{\alpha_s^2}{2}\,
e^{\imath\left(E_{m_s-2}-E_{m_s}\right)\tau/\hbar},\nonumber \\&&
 \varphi^{\left(m_s,m_s\right)}_{m_s,m_s-1}\left(\tau\right)=
\imath \alpha_s e^{\imath\left(E_{m_s-1}-E_{m_s}\right)\tau/\hbar},\hspace{0.5em}
\varphi^{\left(m_s,m_s\right)}_{m_s,m_s}\left(\tau\right)=
-2\alpha_s^2,\nonumber \\ &&
\varphi^{\left(m_s,m_s\right)}_{m_s,m_s+1}\left(\tau\right)
=\imath
\alpha_s \,
e^{\imath\left(E_{m_s+1}-E_{m_s}\right)\tau/\hbar},\hspace{0.5em}
\varphi^{\left(m_s,m_s\right)}_{m_s,m_s+2}\left(\tau\right)=-
\frac{\alpha_s^2}{2}\nonumber \\&& \times\,e^{\imath\left(E_{m_s+2}-E_{m_s}\right)\tau/\hbar},
\hspace{0.5em}
\varphi^{\left(m_s,m_s\right)}_{m_s+1,m_s-1}\left(\tau\right)=
\alpha_s^2 \,
e^{\imath\left(E_{m_s-1}-E_{m_s+1}\right)\tau/\hbar}, \nonumber \\ && \varphi^{\left(m_s,m_s\right)}_{m_s+1,m_s}\left(\tau\right)=-\imath
\alpha_s \,
e^{\imath\left(E_{m_s}-E_{m_s+1}\right)\tau/\hbar},\nonumber\\
 &&\varphi^{\left(m_s,m_s\right)}_{m_s+1,m_s+1}\left(\tau\right)=
\alpha_s^2, \hspace{0.5em}
\varphi^{\left(m_s,m_s\right)}_{m_s+2,m_s}\left(\tau\right)= -\frac{
\alpha_s^2}{2} \,
e^{\imath\left(E_{m_s}-E_{m_s+2}\right)\tau/\hbar}. \nonumber
 \end{eqnarray}
For the sake of shortness, we omit the master equations driving the time evolution of the remaining terms, since the structure is similar to those reported above.

   For times $t\gg 1/\Omega$, the contribution of the
   oscillating terms to the convolution product is negligible and
   the above master equations get simplified forms:
   \begin{eqnarray}
       \dot{ \rho}_{1_L, 1_L}(t)&\simeq& \imath \Omega\left(\rho_{1_L, 1_R}(t)-
\rho_{1_R, 1_L}(t)\right)+ \frac{\alpha_L^2}{2}\int_0^t \Phi
\left(t-t^{\prime}\right)\nonumber \\ &&\times \Big(2\rho_{2_L,
2_L}\left(t^{\prime}\right)-\rho_{1_L,
1_L}\left(t^{\prime}\right)-\rho_{1_R,
1_R}\left(t^{\prime}\right)\Big) dt^{\prime},
   \label{SimplGenMastEq1L1L} \\
\dot{ \rho}_{1_L, 1_R}(t)&\simeq&\imath \Omega\left(\rho_{1_L, 1_L}(t)-
\rho_{1_R, 1_R}(t)\right)- \frac{\alpha_L^2+\alpha_R^2}{4}\nonumber \\&&\times\, \int_0^t \Phi
\left(t-t^{\prime}\right)\Big(\rho_{1_L,
1_R}\left(t^{\prime}\right)+ \rho_{1_R,
1_L}\left(t^{\prime}\right)\Big) dt^{\prime},
\label{SimplGenMastEq1L1R}
\\
\dot{ \rho}_{1_L, 2_L}(t)&\simeq&\imath \left(\omega_{2_L
1_L}\rho_{1_L, 2_L}(t)-\Omega \rho_{1_R, 2_L}(t)\right)+\imath \frac{\alpha_L}{2}
 \int_0^t \Phi \left(t-t^{\prime}\right)\nonumber \\ &&\times\,\Big(\rho_{1_L,
1_L}\left(t^{\prime}\right)+\rho_{1_R,
1_R}\left(t^{\prime}\right)-2\rho_{2_L,
2_L}\left(t^{\prime}\right)\Big) dt^{\prime}, \label{SimplGenMastEq1L2LD0}
\end{eqnarray}
\begin{eqnarray}
\dot{ \rho}_{1_L, 2_R}(t)&\simeq&\imath \left( \omega_{2_R 1_R} \rho_{1_L, 2_R}(t)
-\Omega \rho_{1_R, 2_R}(t)\right)+\imath \frac{\alpha_R}{2}\nonumber \\&& \times \, \int_0^t
\Phi \left(t-t^{\prime}\right)\Big(\rho_{1_L,
1_R}\left(t^{\prime}\right)+
\rho_{1_R,
1_L}\left(t^{\prime}\right)
\Big), \,\,\,\,\label{SimplGenMastEq1L2RD0}
\\
\dot{ \rho}_{2_L, 2_L}(t)&\simeq& \frac{\alpha_L^2}{2} \int_0^t
\Phi\left(t-t^{\prime}\right)\Big( \rho_{1_L,
1_L}\left(t^{\prime}\right)+\rho_{1_R, 1_R}\left(t^{\prime}\right)-
4\rho_{2_L, 2_L}\left(t^{\prime}\right)\nonumber \\ &&+2 \rho_{3_L,
3_L}\left(t^{\prime}\right)\Big) dt^{\prime},
\label{SimplGenMastEq2L2LD0}
\\
  \dot{ \rho}_{m_s, m_s}(t) &\simeq&\alpha_s^2
\int_0^t\Phi\left(t-t^{\prime}\right)
\Big(\rho_{m_{s-1},m_{s-1}}\left(t^{\prime}\right)-2
\rho_{m_{s},m_{s}}\left(t^{\prime}\right)\nonumber \\&&+
\rho_{m_{s+1},m_{s+1}}\left(t^{\prime}\right) \Big)d t^{\prime},\hspace{0.5em}
 m=3,\ldots,N_s-1, \hspace{0.5em} s=L,R,\label{SimplGenMastEqmsmsD0}
\\
\dot{ \rho}_{N_{s},N_{s}}(t)&\simeq &\alpha_s^2\int_0^t
\Phi\left(t-t^{\prime}\right)\Big(
\rho_{N_{s}-1,N_{s}-1}\left(t^{\prime}\right)
-\rho_{N_s,N_s}\left(t^{\prime}\right)
\Big)dt^{\prime}, \label{SimplGenMastEqNsNsD0}\\&& s=L,R. \nonumber
\end{eqnarray}
The dynamics of the populations,
$\rho_{m_s, m_s}(t)$, and the coherent term, $\rho^c(t)$, results to be
decoupled from the time evolution of remaining
coherence terms, $\rho_{m_s, n_{s^{\prime}}}(t)$, for every
$m_s, n_{s^{\prime}}\neq1$ and $s,s^{\prime}=L,R$,
undergoing "fast" oscillations \cite{Zeno0} and
giving vanishing contribution for $t\gg 1/\Omega$.
The above equations are recovered through the adiabatic theorem \cite{AT}, we omit the calculations for the sake of shortness.

\section{The adiabatic theorem}\label{SAT}

We now prove the consistency of the master equations obtained in the previous Appendix
through the adiabatic theorem described in Ref. \cite{AT}. Since the dynamics of the eigenspace of the operator $H_0$, appearing in Eq. (\ref{Lf}), is closed, by defining the reduced operator
\begin{eqnarray}
&&\hspace{-4em}\Pi\left[A\right]=\bar{A}=Q A Q+ \sum_{s=L,R}\sum_{n_s=2}^{N_s}
P_{n_s} A P_{n_s}, \hspace{1em} Q=P_{1_+}+
P_{1_-},\nonumber
\end{eqnarray}
the relationships
\begin{eqnarray}
 &&R\left(t,t^{\prime}\right)= e^{\mathcal{L}_0
\left(t-t^{\prime}\right)}\left[\rho\left(t^{\prime}\right)\right],
\hspace{1em}P_{1_{\pm}}=\left|1_{\pm}\rangle\langle
1_{\pm}\right|,\nonumber \\ &&P_{n_s}=\left|n_s\rangle\langle n_s\right|,
 \hspace{1em} n_s=2,\ldots,N_s,\hspace{1em}s=L,R,\hspace{1em}\bar{V}=0, \nonumber\\
  &&\Pi\left[V^2\right]=\sum_{s=L,R} \alpha_s^2
\left(P_{1_s} + P_{N_s}+2\sum_{n_s=2}^{N_s-1}
P_{n_s}\right), \nonumber \\
&&\alpha_L^2 P_{1_L}+\alpha_R^2
P_{1_R}=p_{1_+,1_+} \left|1_+\rangle \langle 1_+\right|+p_{1_-,1_-}
\left|1_-\rangle \langle 1_- \right|+ p_{1_-,1_+} \left|1_- \rangle \langle 1_+\right|\nonumber \\ &&
+\,p_{1_+,1_-}
\left|1_+ \rangle \langle 1_-
\right|,\hspace{1em} p_{1_{\pm},1_{\pm}}=\alpha_L^2
\left(\gamma^{\left(\pm\right)}_L\right)^2+
\alpha_R^2\left(\gamma^{\left(\pm\right)}_R\right)^2,\nonumber
\end{eqnarray}
\begin{eqnarray}&&\hspace{-14em}
p_{1_{\pm},1_{\mp}}=\alpha_L^2\gamma^{(+)}_L \gamma^{(-)}_L+
\alpha_R^2 \gamma^{(+)}_R \gamma^{(-)}_R. \nonumber
\end{eqnarray}
lead to the following reduced master equation
\begin{eqnarray}
&&\hspace{-6em}\dot{\bar{\rho}}(t) \simeq
-\frac{\imath}{\hbar}\left[\bar{H},\bar{\rho}(t)\right]+ \int_0^t
\Phi\left(t-t^{\prime}\right)\Big\{-\imath
\left[\bar{V},\bar{R}\left(t,t^{\prime}\right)\right]\nonumber \\&&\hspace{-3em}-\frac{1}{2}
\left\{\Pi\left[V^2\right],\bar{R}\left(t,t^{\prime}\right)\right\}+
\Pi \left[V \bar{R}\left(t,t^{\prime}\right)V\right]
\Big\}\,dt^{\prime},
 \label{GenMastEqAdiab}
\end{eqnarray}
describing the dynamics of the populations of the energy levels.
For long timescales, $t \gg 1/\Omega$, the
oscillating terms give a negligible contribution
to the convolution and we get the following relations:
\begin{eqnarray}
&&-\frac{1}{2}\left\{\Pi\left[V^2\right],\bar{R}\right\}=
-\frac{\alpha_L^2+\alpha_R^2}{2}
\left(\rho_{1_+,1_+}\left(t^{\prime}\right)\left|1_+\rangle \langle
1_+\right|+\rho_{1_-,1_-}\left(t^{\prime}\right)\left|1_-\rangle
\langle 1_-\right|\right)\nonumber \\ &&+\,
\frac{\alpha_R^2-\alpha_L^2}{4}
\left(\rho_{1_+,1_+}\left(t^{\prime}\right)+\rho_{1_-,1_-}
\left(t^{\prime}\right)\right) \,\left(\left|1_+\rangle \langle
1_-\right|+\left|1_-\rangle \langle
1_+\right|\right)\nonumber \\ &&-\sum_{s=L,R}\alpha_s^2\Big\{2
\sum_{n_s=2}^{N_s-1}\rho_{n_s,n_s}\left|n_s\rangle \langle
n_s\right|+\rho_{N_s,N_s}\left|N_s\rangle \langle
N_s\right|\Big\},\nonumber
\\
&&\Pi\left[V
\bar{R}\left(t,t^{\prime}\right)V\right]=\sum_{s=L,R}\alpha_s^2
\Big\{\rho_{2_s,2_s}\left(t^{\prime}\right)\left|1_s \rangle \langle 1_s\right|+\rho_{N_s-1,N_s-1}\left(t^{\prime}\right) \left|N_s \rangle \langle
N_s\right|\nonumber  \\ &&+\left(R_{1_s,1_s}\left(t,t^{\prime}\right)
+\rho_{3_s,3_s}\left(t^{\prime}\right)\right)\left|2_s \rangle
\langle 2_s\right|+
\sum_{n_s=3}^{N_s-1}\big(\rho_{n_s-1,n_s-1}\left(t^{\prime}\right)\nonumber \\ &&+\,
\rho_{n_s+1,n_s+1}\left(t^{\prime}\right) \big)\left|n_s \rangle
\langle n_s\right| \Big\},\nonumber
\end{eqnarray}
where
\begin{eqnarray}
&&\hspace{-4em}R_{1_{L_{\left(R\right)}},1_{L_{\left(R\right)}}}
 \left(t,t^{\prime}\right)=
 \left(\gamma^{\left(+\right)}_{L_{\left(R\right)}}\right)^2
\rho_{1_+,1_+}\left(t^{\prime}\right) +\,
\left(\gamma^{\left(-\right)}_{L_{\left(R\right)}} \right)^2
\rho_{1_-,1_-}\left(t^{\prime}\right)\nonumber \\&&\hspace{-4em}+
\gamma^{\left(+\right)}_{L_{\left(R\right)}}
\gamma^{\left(-\right)}_{L_{\left(R\right)}}\Big(e^{-\imath
\omega_{1_+,1_-}\left(t-t^{\prime}\right)}
\rho_{1_+,1_-}\left(t^{\prime}\right)+\,e^{\imath
\omega_{1_+,1_-}\left(t-t^{\prime}\right)}
\rho_{1_-,1_+}\left(t^{\prime}\right)\Big), \nonumber
\end{eqnarray}
giving the following master equations
\begin{eqnarray}
\dot{ \rho}_{1_{\pm},1_{\pm}}(t)&\simeq& \int_{0}^t
\Phi\left(t-t^{\prime}\right)\Big\{ \alpha_L^2
\left(\gamma_L^{\left({\pm}\right)}\right)^2
\rho_{2_L,2_L}\left(t^{\prime}\right) +\alpha_R^2
\left(\gamma_R^{\left({\pm}\right)}\right)^2
\rho_{2_R,2_R}\left(t^{\prime}\right)\nonumber \\ &&-\,\frac{\alpha_L^2+
\alpha_R^2}{2}\,\rho_{1_{\pm},1_{\pm}}\left(t^{\prime}\right) \Big\}\,
dt^{\prime},
\nonumber
\end{eqnarray}
\begin{eqnarray}
\dot{ \rho}_{1_{\pm},1_{\mp}}(t)&\simeq&
\mp \,2 \imath \Omega \rho_{1_{\pm},1_{\mp}} (t) +\int_{0}^t \Phi\left(t-t^{\prime}\right)\Bigg\{
\frac{\alpha_R^2-\alpha_L^2}{4}
\Big(\rho_{1_{+},1_{+}}\left(t^{\prime}\right)\nonumber \\ &&+\,
\rho_{1_{-},1_{-}}\left(t^{\prime}\right)\Big)+ \alpha_L^2
\gamma_L^{\left(+\right)}
\gamma_L^{\left(-\right)}\rho_{2_L,2_L}\left(t^{\prime}\right)\nonumber \\&&
+\,\alpha_R^2 \gamma_R^{\left(+\right)} \gamma_R^{\left(-\right)}
\rho_{2_R,2_R}\left(t^{\prime}\right) \Bigg\} \, dt^{\prime}.
\nonumber
\end{eqnarray}

We are now equipped to build up the
master equations generated by
Eq. (\ref{GenMastEqAdiab}). After some
long but straightforward algebra we recover the \emph{same}
equations for the dynamics of the levels populations and
off diagonal elements of the density matrix, given by Eqs.
(\ref{SimplGenMastEq1L1L}), (\ref{SimplGenMastEq1L1R}), (
\ref{SimplGenMastEq1L2LD0}), (\ref{SimplGenMastEq1L2RD0}),
(\ref{SimplGenMastEq2L2LD0}), (\ref{SimplGenMastEqmsmsD0}),
(\ref{SimplGenMastEqNsNsD0}), obtained in the previous Appendix.

\section{Coherence and populations}\label{B}

This Appendix is devoted to a detailed analysis of the populations and the coherence terms through their Laplace transforms, useful to recover the dynamics over long time scales.

We consider Eqs. (\ref{SimplGenMastEq1L1L}), (\ref{SimplGenMastEq1L1R}), (\ref{SimplGenMastEq2L2LD0}), (\ref{SimplGenMastEqmsmsD0}), (\ref{SimplGenMastEqNsNsD0}) and those obtained by exchanging the indexes $L$ and $R$, the corresponding Laplace transforms show an iterative structure of continued fraction difficult to inverted, thus, in order to evaluate $p_{1_L}(t)$, $p_{1_R}(t)$ and $p^c(t)$, the reasonable assumption that the number of internal energy levels is infinite. In this way, Eq. (\ref{SimplGenMastEqmsmsD0}) gives the following difference equation:
\begin{eqnarray}
&&\tilde{p}_{n_s+2}(u)-2 \left(\frac{u}{2 \alpha_s^2
\tilde{\Phi}(u)}+1\right)\tilde{p}_{n_s+1}(u)+\tilde{p}_{n_s}(u)=0,\nonumber
\\&&
n_s=2,\ldots,N_s\hspace{2em} s=L,R. \nonumber
\end{eqnarray}
Following the arguments reported in the last paragraph of Section \ref{2}, the solutions of the above equation are approximated by the solutions of the second order difference equation obtained by setting $N_s=\infty$, for $s=L,R$. In this way, the following expressions are obtained:
\begin{eqnarray}
&&\hspace{-1.8em}\tilde{p}_{n_s}(u)=A_s(u)\left(\lambda^{(s)}_+(u)
\right)^{n_s}+B_s(u)\left(\lambda^{(s)}_-(u)\right)^{n_s}, \hspace{0.6em}n_s=2,\ldots,N_s\hspace{0.6em} s=L,R, \nonumber
\end{eqnarray}
where $A_s(u)$ and $B_s(u)$ are independent of $n_s$ and
\begin{eqnarray}
&&\hspace{-1.8em}\lambda^{(s)}_{\pm}(u)=1+\frac{u }{ 2
\alpha_s^2 \tilde{\Phi}(u)}\pm\sqrt{\left(1+\frac{u }{ 2 \alpha_s^2
\tilde{\Phi}(u)}\right)^2-1}, \hspace{0.6em}
n_s=2,3,\ldots, \hspace{0.6em}s=L,R. \nonumber
\end{eqnarray}
The Laplace transform of normalization constraint of the whole levels populations, $\sum_{s=L,R}\sum_{n_s=1}^{\infty} p_{n_s}(t)=1$, reads $\sum_{s=L,R}\sum_{n_s=1}^{\infty}\tilde{p}_{n_s}(u)=1/u$. Also, the functions $\tilde{\Phi}(u)$ analyzed are positive for every $u>0$ and $s=L,R$, thus, $\lambda^{(s)}_+(u)>1$ for every $u>0$ and $s=L,R$. Furthermore, the convergence of the series
$\sum_{s=L,R}\sum_{n_s=2}^{\infty}  A_s(u)\left(\lambda^{(s)}_+(u)\right)^{n_s}$
gives $ A_s(u)=0$ for every $u>0$ and $s=L,R$, and it results
\begin{equation}
\hspace{-3.5em}\tilde{p}_{n_s}(u)=B_s(u) \left(\lambda^{(s)}_-(u)\right)^{n_s}, \quad
n_s=2,3,\ldots,\quad s=L,R. \nonumber
\end{equation}
The function $B_s(u)$ is fixed by Eq. (\ref{SimplGenMastEq2L2LD0}) and the one obtained by the mutual exchange of $L$ and $R$,
\begin{equation}
\hspace{-5.5em}B_s(u)=\frac{-\alpha_s^2
\tilde{\Phi}(u)\left(\tilde{p}_{1_L}(u)+\tilde{p}_{1_R}(u)\right)}{
2 \left(\lambda_-^{(s)}(u)\right)^2\left(\alpha_s^2
\left(\lambda_-^{(s)}(u)-2\right)
\tilde{\Phi}(u)-u\right)}, \nonumber
\end{equation} for every $s=L,R$.
The dynamics of the populations of the lower energy levels,
$p_{1_L}(u)$, $p_{1_R}(u)$ and the coherence term, $p^c(u)$,
is describe by the matrix form $\mathbf{M}(u)\cdot \mathbf{X}(u)=\mathbf{Y}$, where $\mathbf{M}(u)$, $\mathbf{X}(u)$ and $\mathbf{Y}$ read
\begin{eqnarray}
&&\hspace{-1em}\left[\mathbf{M}(u)\right]_{1,1}=u+
\frac{\alpha_L^2}{2}\tilde{\Phi}(u) + \frac{\alpha_L^4
\left(\tilde{\Phi}(u)\right)^2}{2\left( \alpha_L^2
\tilde{\Phi}(u)\left(\lambda_-^{(L)}(u)-2\right)-u\right)},\nonumber \\&&\hspace{-1em}
\left[\mathbf{M}(u)\right]_{1,2}=\frac{\alpha_L^2}{2}
\tilde{\Phi}(u)\left(1+
 \frac{\alpha_L^2\tilde{\Phi}(u)}{\alpha_L^2\tilde{\Phi}(u)
\left(\lambda_-^{(L)}(u)-2\right)-u}\right), \nonumber \\
&&\hspace{-1em}\left[\mathbf{M}(u)\right]_{1,4}=-\left[\mathbf{M}(u)\right]_{1,3}=
\left[\mathbf{M}(u)\right]_{2,3}=-\left[\mathbf{M}(u)\right]_{2,4}=
-\left[\mathbf{M}(u)\right]_{3,1}\nonumber \\ &&\hspace{-1em}=\left[\mathbf{M}(u)\right]_{3,2}=
\left[\mathbf{M}(u)\right]_{4,1}=-\left[\mathbf{M}(u)\right]_{4,2}=
 \imath \Omega,\nonumber \\&&\hspace{-1em}\left[\mathbf{M}(u)\right]_{2,1}=\frac{\alpha_R^2}{2}
\tilde{\Phi}(u)\left(1+
 \frac{\alpha_R^2\tilde{\Phi}(u)}{\alpha_R^2\tilde{\Phi}(u)
\left(\lambda_-^{(R)}(u)-2\right)-u}\right), \nonumber
\\
&&\hspace{-1em}
\left[\mathbf{M}(u)\right]_{2,2}=u+
\frac{\alpha_R^2}{2}\tilde{\Phi}(u) + \frac{\alpha_R^4
\left(\tilde{\Phi}(u)\right)^2}{2\left(
\alpha_R^2\tilde{\Phi}(u)\left(\lambda_-^{(R)}(u)-2\right)-u\right)},
\nonumber
\\
&&\hspace{-1em}\left[\mathbf{M}(u)\right]_{3,3}=\left[\mathbf{M}(u)\right]_{4,4}=
\frac{\alpha_L^2+\alpha_R^2}{4}\tilde{\Phi}(u)+u,\nonumber
\\
&&\hspace{-1em}
\left[\mathbf{M}(u)\right]_{3,4}=\left[\mathbf{M}(u)\right]_{4,3}=
\frac{\alpha_L^2+\alpha_R^2}{4}\tilde{\Phi}(u),\nonumber \\
&&\hspace{-1em}\mathbf{X}(u)=\left[
\begin{array}{c}
\tilde{p}_{1_L}(u) \\ \tilde{p}_{1_R}(u) \\
\tilde{\rho}_{1_L, 1_R}(u) \\ \tilde{\rho}_{1_R, 1_L}(u)
\end{array}
\right],\hspace{0.5em} \mathbf{Y}=\left[
\begin{array}{c}
1 \\0 \\0 \\ 0
\end{array}
\right]. \nonumber
\end{eqnarray}
Finally, the Laplace transform of the coherence term results to be
\begin{eqnarray}
\tilde{p}^c(u)&=&-4 \Omega \left(u+2 \alpha_L^2 \tilde{\Phi}(u) +
u^{1/2}F_L(u) \right)\Big(u^{3/2}+u F_R(u)+ \alpha_R^2
\tilde{\Phi}(u) \nonumber \\ &&\times \,\left(3 u^{1/2}+F_R(u)\right)
\Big)\Bigg/\Bigg(\left(u^2+4 \Omega^2\right)\Bigg(u^{1/2}\left(u^{1/2}+F_L(u)\right)\nonumber \\
&&\times
 \Big(2 u \left(u^{1/2}+F_R(u)\right)+\alpha_R^2 \tilde{\Phi}(u)
\left(5u^{1/2}+F_R(u)\right) \Big) +\alpha_L^2 \tilde{\Phi}(u)\nonumber \\ &&\times \,\Big(
u^{1/2}\left(5u^{1/2}+F_L(u)\right) \left(u^{1/2}+F_R(u)\right)+2 \alpha_R^2\tilde{\Phi}(u)\nonumber\\
&&\times  \left(6
u^{1/2}+F_L(u)+F_R(u)\right)
\Big)\Bigg),\label{pc} \\
&&F_s(u)=\sqrt{u+4\alpha_s^2
\tilde{\Phi}(u)}, \hspace{1em} s=L,R, \nonumber
\end{eqnarray}
while the population of the ground level $L$ reads
\begin{eqnarray}
\tilde{p}_{1_L}(u)&=&\left(u+2 \alpha_L^2 \tilde{\Phi}(u)+u^{1/2}F_L(u)\right)\Bigg(2 u^{1/2}\left(u^2+2 \Omega^2\right)\left(u^{1/2}+F_R(u)\right)\nonumber \\ &&+\,
\alpha_R^2\tilde{\Phi}(u)\left(8 \Omega^2+5 u^2+u^{3/2}F_R(u)\right)\Bigg)\Bigg/
\Bigg(u^{1/2}\left(u^2+4 \Omega^2\right)\Big(u^{1/2}\nonumber\\
&&\left(u^{1/2}+F_L(u)\right)
\left(2 u\left(u^{1/2}+F_R(u)\right)+\alpha_R^2 \tilde{\Phi}(u)\left(5 u^{1/2}+F_R(u)\right)\right) \nonumber\\
&&+\,\alpha_L^2 \tilde{\Phi}(u)\big(u^{1/2}\left(5 u^{1/2}+F_L(u)\right)\left(u^{1/2}+F_R(u)\right)+
2 \alpha_R^2 \tilde{\Phi}(u)\nonumber \\ && \times \,\left(6 u^{1/2}+F_L(u)+F_R(u)\right)\big)\Big)\Bigg).\label{P1Lu}
\end{eqnarray}
The expression for $\tilde{p}_{1_R}(u)$ is obtained from the above relation through the mutual exchange of the indexes $L$ and $R$. Notice that the equality (\ref{FinalEqs2}) is fulfilled.

 The inverse power law behavior of the coherence term, described by Eq. (\ref{pc}), is evaluated for various expressions of the memory kernel $\hat{\Phi}(u)$, through the converging term by term inverse Laplace transform of the corresponding series expansion around $u=0$, since such series converges in a sector $\left|\arg u\right|\leq \chi_0$, where $\pi/2< \chi_0 \leq \pi$. The choice of the long time scales derives from the condition of convergence of the related series expansions.

 \section{Acknowledgments}
This work is based upon research supported by the South African Research Chairs Initiative of
the Department of Science and Technology and National Research Foundation.
F.G. is grateful to Prof. P. Facchi for the fruitful suggestions on the model and to Dr. A.A. Budini  for the useful discussions on the CTQRW.


\begin{thebibliography}{91}
\bibitem{BP} H.P. Breuer and F. Petruccione, \emph{The Theory of Open Quantum Systems},
 Oxford University Press, Oxford (2002).
\bibitem{AL} R. Alicki and K. Lendi, \emph{Quantum Dynamical Semigroups and Applications}, Lecture Notes in Physics Vol. {\bf 286}, Springer, Berlin (1987).
    \bibitem{GKS76} V. Gorini, A. Kossakowski and E.C.G. Sudarshan, J. Math. Phys. {\bf 17}, 821 (1976).
    \bibitem{Lindblad76} G. Lindblad, Comm. Math. Phys. {\bf 48}, 119 (1976).
\bibitem{MW} E.W. Montroll and G.H. Weiss, J. Math. Phys. {\bf 6}, 167 (1965).
\bibitem{CTQRW0} A. A. Budini, Phys. Rev. A {\bf 69}, 042107 (2004).
\bibitem{budiniPRE04} A. A. Budini, Phys. Rev. E {\bf 72}, 056106 (2005).

\bibitem{S} I.M. Sokolov, Phys. Rev E {\bf 66}, 041101 (2002).
\bibitem{Zeno0} D. Bruno, P. Facchi, S. Longo, P. Minelli, S.
Pascazio, and A. Scardicchio, J. Phys. Chem. A, {\bf 113}, 14875 (2009).





\bibitem{MS} A. Beskow and J. Nilsson, Ark. Fys. {\bf 34}, 561 (1967); L.A. Khalfin, JETP Lett. {\bf 8}, 65 (1968); B. Misra and E.C.G. Sudarshan, J. Math Phys. {\bf 18},
756 (1977).


\bibitem{Cook} R.J. Cook, Phys. Scr. T {\bf 21}, 49 (1988).



\bibitem{IHBW} W.M. Itano, D.J. Heinzen, J.J. Bollinger and D.J. Wineland,
Phys. Rev. A {\bf  41}, 2295 (1990).

\bibitem{FGR} M.C. Fisher, B. Gutierrez-Medina and M.G. Rainez,
Phys. Rev. Lett. { \bf 87}, 040402 (2001).


\bibitem{HW} D. Home and M.A.B. Whitaker, Ann. Phys. (N.Y.) {\bf 258},
237 (1997).



\bibitem{BN} A. Beskow and J. Nilsson, Ark. Fys. {\bf 34},
561 (1967).
\bibitem{K} L.A. Khalfin, JETP Lett. {\bf 8}, 65 (1968).
\bibitem{FNPPRL} P. Facchi, H. Nakazano and S. Pascazio, Phys. Rev. Lett. {\bf 86}, 2699 (2001).
    \bibitem{P} A.D. Panov, Phys. Lett. A {\bf 298}, 295 (2002).
\bibitem{CH3F}B. Nagels, L.J.F. Hermans  and P.L. Chapovsky, Phys. Rev. Lett. {\bf 79}, 3097 (1997).


        \bibitem{NNP} H. Nakazato, M. Namiki and S. Pascazio, Int. J. Mod. Phys. B {\bf 10}, 247 (1996).

    \bibitem{BBGG}M. Bologna, A. A. Budini, F. Giraldi and P. Grigolini,
Journ. Chem. Phys. {\bf 130}, 244106 (2009).


\bibitem{CTQRW1} A. A. Budini and P. Grigolini, Phys. Rev. A {\bf 80},
022103 (2009)


\bibitem{widder} D.V. Widder, \emph{The Laplace Transform}, Princeton Univ. Press, Princeton, NJ (1941).
    \bibitem{Mainardi} R. Gorenflo and F. Mainardi, \emph{Integral and Differential Equations of Fractional Order}, CISM LECTURE NOTES, Bologna, (2000).
        \bibitem{MK} R. Metzler and J. Klafter, Phys. Rep {\bf 339}, 1 (2000).
        \bibitem{prabakar} T.R.  Prabhakar, Yokohama Mathematical Journal {\bf19}, 7 (1971).
            \bibitem{MainardiBook} F. Mainardi, \emph{Fractional Calcolus and Waves in Linear Viscoelasticity}, Imperial College Press, World Scientific Publishing, (2010).
\bibitem{ZK} G. Zumofen and J. Klafter, Phys. Rev. E {\bf 47}, 851 (1993).
\bibitem{MSacp1958}E.W. Montroll and K.E. Shuler, Adv. Chem. Phys. {\bf 1}, 361 (1958).
\bibitem{CM1980}M. Capitelli and E. Molinari, \emph{Kinetic of Dissociation Process in Plasma in the low and Intermediate Pressure Range.} Topics in Current Chemistry, Springer, Vol. {\bf 90}, New York (1980).
\bibitem{AT} P. Facchi and S. Pascazio, Phys. Rev. Lett. {\bf 89}, 080401 (2002);  J. Phys. Soc. Japan. {\bf 72} Supplement C, 30 (2003).

\end{thebibliography}
\end{document}